\documentclass[12pt]{article}
\pdfoutput=1
\usepackage{slashed}
\usepackage{amsmath,amssymb}
\usepackage{bbm}
\usepackage{subfigure}

\usepackage{hyperref}
\usepackage{graphicx}
\usepackage{amsbsy}

\begin{document}
\setlength{\unitlength}{1mm}
\begin{titlepage}
\vspace*{0.8cm}
\begin{center}
 {\huge \bf  Slinky Evolution of Domain Wall Brane Cosmology }
\end{center}
\vskip0.2cm

\begin{center}
 {\bf Avihay Kadosh$^a$, Aharon Davidson$^b$ and Elisabetta Pallante$^a$}
\end{center}
\vskip 4pt

\begin{center}
$^a$ {\it Centre for Theoretical Physics, University of Groningen,
9747 AG, Netherlands}\\
$^b$ {\it Physics Department, Ben-Gurion University of the Negev,
Beer-Sheva 84105 Israel}

\vspace*{0.1cm}

{\tt a.kadosh@rug.nl, davidson@bgu.ac.il, e.pallante@rug.nl }
\end{center}
\vglue 0.3truecm

\begin{abstract}
\vskip 3pt \noindent

Invoking an initial symmetry between the time $t$ and some extra
spatial dimension $y$, we discuss a novel scenario where the
dynamical formation of the 4-dim brane and its cosmological
evolution are induced simultaneously by a common $t
\leftrightarrow y$ symmetry breaking mechanism. The local maximum
of the underlying scalar potential is mapped onto a `watershed'
curve in the $(t,y)$ plane; the direction tangent to this curve is
identified as the cosmic time, whereas the perpendicular direction
serves locally as the extra spatial dimension. Special attention
is devoted to the so-called slinky configurations, whose brane
cosmology is characterized by a decaying cosmological constant
along the watershed curve. Such a slinky solution is first
constructed within a simplified case where the watershed is
constrained by $y=0$. The physical requirements  for a slinky
configuration to generate a realistic model of cosmological
evolution are then discussed in a more elaborated framework.

\end{abstract}
\end{titlepage}

\section{Introduction}

Extra dimensional theories of particle physics and cosmology have
received widespread attention  for more than a decade now. As the typical scale of new
physics in phenomenologically plausible extra
dimensional scenarios is naturally around a TeV, we hope to be
able to test them  at the Large Hadron Collider (LHC) and other
contemporary experiments. The original idea dates back to the works
of Kaluza and Klein (KK) \cite{Kaluza:1921tu,Klein:1926tv} aiming
at a unified theory of electromagnetism and gravity. In the KK
construction \cite{Klein:1926tv} all known matter fields propagate in the full extra
dimensional spacetime and the 4D low energy effective theory is
obtained by compactifying the extra dimension on a circle and keeping the zero mode of the KK tower.
More recently, string theory constructions including D-branes and
in particular the Horava-Witten solution of supergravity
\cite{HoravaWitten1,HoravaWitten2} inspired Arkani-Hamed,
Dimopolous and Dvali (ADD) to introduce the so called {\it Large
Extra Dimensions} paradigm in order to solve the gauge hierarchy
problem \cite{ADD1,ADD2,ADD3}. In the ADD model, only gravity is
free to propagate in the extra dimensional space, while SM
fields are confined to a $3+1$ dimensional hypersurface, referred
to as a {\it brane}. The ADD construction allows, in principle,
for ``large" extra dimensions already at a $mm$ scale,
corresponding to a very small 5D Planck mass. However, recent
experiments probing the short distance behavior of Newtonian
inverse square law have already placed upper bounds of
$\mathcal{O}(40\mu m)$ on the size of the extra dimensions for
relevant realizations of the ADD idea, see for example
\cite{SRGravity1:Kapner:2006si,SRGravity2:Geraci:2008hb,SRGravity3:Sushkov:2011zz}.
Shortly after, Randall and Sundrum (RS) \cite{RS1} offered an
alternative to compactification in the form of a warped extra
dimension confined between two branes, referred to as the UV and
IR brane. In the RS construction, it is again only gravity which
propagates in the full 5D space time, while the SM fields are
confined to the IR brane. The warped geometry implies a varying 5D
scale along the extra dimension which is of $\mathcal{O}(M_{Pl})$
at the UV brane and of $\mathcal{O}({\rm TeV})$ at the IR brane.
By pushing the
IR brane to infinity, it has been shown \cite{RS2} that it is possible
to localize gravity to the UV brane and that deviations from the
Newtonian inverse square law become relevant at very small length
scales of $\mathcal{O}({\rm TeV})^{-1}$.

In  the ADD, RS and some previous constructions
\cite{Visser:1985qm,GibbonsBrane86} the branes are fundamental,
namely they are treated as infinitely thin delta distribution
sources. The 4D effective theory is obtained by performing a KK
decomposition of brane and bulk fields and studying the
corresponding interactions. Main implications
of these constructions are new KK particles with masses of
$\mathcal{O}(TeV)$ and modifications of the Newtonian potential,
both of which are already being challenged by  LHC results and
short range gravity experiments \cite{SRGravity1:Kapner:2006si}.
Depending on the  theoretical setup, additional experimental constraints
can come from precision measurements of rare decays
\cite{Agashe:2004cp}, electroweak precision observables
\cite{CarenaEW2003}, the Cosmic Microwave Background (CMB)
anisotropies \cite{BraxCosmologyRev.}  and more.

Subsequently, the original extra dimensional setups were
modified and extended in the context of both particle physics and
cosmology. The cosmological evolution of fundamental branes was
already studied shortly after the introduction of the RS model
\cite{BraneCosmology1:Binetruy:1999ut,BraneCosmology2:Csaki:1999jh,BraneCosmology3:Cline:1999ts,
BraneCosmology4:Chung:1999zs,BraneCosmology5:Binetruy:1999hy,BraneCosmology6:Flanagan:1999cu}
, exhibiting a non standard cosmological evolution above an energy
scale related to the brane tension and the curvature of the bulk.
 In the last  decade, various
aspects of brane world cosmology have been extensively studied
including inflation, gravitational perturbations and late time
acceleration,  with emphasis on possible signatures in the CMB
anisotropy spectrum \cite{BraxCosmologyRev.}. More exotic models
considered the possibility of colliding branes, with fundamental
branes
\cite{CollidingBranes:Dvali:1998,EkpyroticSteinhardt:2001,CollidingBranes:Takamizu:2004rq}
or within the $D3-D7$ string theory inspired constructions
\cite{D3D7Baumann:2007}.

If our universe is indeed a 3-{\it brane}, coming from a string
theory realization of RS or other braneworld setups
\cite{RSstring:Chan:2000ms,RSstring:Altshuler:2005ca,RSstring:Acharya:2006mx},
treating it as an infinitely thin brane suffices  for deriving
the 4D particle physics and cosmology. Effects of finite thickness of the brane can  be
accounted for by averaging procedures developed in
\cite{ThickBrane1:Kanti:1999sz,ThickBrane2:Mounaix:2002mm,ThickBrane3:Santiago:2005uq,ThickBrane4:Cvetic:2008gu}.

Alternatively to string theory, the simplest underlying dynamics that generates a brane
may have a field theoretical origin; the
brane would correspond to a topological defect (domain wall (DW),
vortex, etc.)  in higher dimensions. The idea that our universe is
a DW brane in higher dimensional space time dates back to the work of
Rubakov and Shaposhnikov \cite{DWRubakov:1983bb} \footnote{Higher
dimensional topological defects were also considered
around the same time \cite{VortexAkama:1982jy}, but are not the
focus of the present work.}.  DW configurations can be
generated by a scalar field supported by a double well or periodic
potential and are able to provide a smooth realization
of the brane and the RS warped metric
\cite{DeWolfe,Gremm:Dw:1999pj,Gremm:dS4:2000dj,Davidson,GiovanniniM4}.
In these setups the width of the DW in the extra dimension
plays an important role, in particular in the localization
mechanisms of gravity, gauge and matter fields to the brane
\cite{DwGravity:Csaki:2000fc,Kehagias(Localization):2000au,DwGravityNum:Bazeia:2007nd}.

What is common to all of these constructions is the maximally 4
symmetric geometry associated with the DW brane, allowing for a
Minkowski ($M_4$), $dS_4$ and $AdS_4$ brane solutions. The $dS_4$
case already introduces difficulties in the form of (naked)
curvature singularities  at a finite distance from the DW brane
\cite{Gremm:dS4:2000dj,Davidson}, which can be overcome by
compensating divergences in the energy momentum tensor of the DW
scalar \cite{Davidson}. A  $dS_4$ DW solution where the
singularities are relaxed to be horizons can  be obtained
numerically \cite{Gremm:dS4:2000dj,Volkas-Slatyer:2006un}; however,
it still introduces problems for the localization of fermions and
possibly other fields.

Differently from the maximally 4-symmetric cases,  there has
been so far  only one attempt at finding  maximally 3-symmetric DW
configurations, the latter being relevant for cosmology
\cite{Giovannini}. The resulting solutions are ``time shifted"
versions of the maximally 4-symmetric solutions of
\cite{GiovanniniM4},  yielding a bouncing cosmological evolution
opposite to the desired one:  $H(\tau>0)<0$ and
$H(\tau<0)>0$, where $\tau$ denotes a conformal time coordinate,
related to the cosmological time $t$ by $a(\tau)d\tau=t$
\cite{Giovannini}.

The effective cosmology of fermionic and scalar fields in the
vicinity of an unspecified time dependent DW configuration in 5D
was studied recently in \cite{DamienTime}, as a sequel to previous
attempts to obtain the Standard Model (SM) of particle physics
as the low energy effective theory on a Minkowski DW brane in 5D
\cite{DamienSM}. In this setting electroweak symmetry breaking
(EWSB) is driven by the DW scalar itself. A subtle interplay
between the bulk scalar fields, the DW scalar and gauge fields
is needed in order for the SM particle content (Higgs
and fermions) and the SM symmetry group to be confined to the
brane. This construction was later extended to a larger Grand
Unified Theory (GUT) symmetry group in \cite{DavidsonE6}.
 For a thorough overview of the
various aspects of DW configurations (Thick Brane solutions) the
interested reader is referred to \cite{DWReview}.

In this paper we aim at obtaining cosmologically plausible time
dependent DW configurations, following a rather different approach
from the above attempts. This tells us that we have to adopt a
Friedmann-Lema$\hat{i}$tre-Robertson-Walker (FLRW), maximally 3-symmetric
ansatz for the 5D geometry, with crucial implications as will be explained below.

 The paper is organized as follows. In
 Sec.~\ref{Sec:definiton+solution} we elaborate on our approach to
 the problem and define the notion of ``slinky"  DW
 configurations. In Sec.~\ref{Sec:Solution} we  obtain the
 simplest solution with most of the desirable characteristics described in
 Sec.~\ref{Sec:definiton+solution}, and discuss the problematic features of
this first attempt. In Sec.~\ref{Sec:DWtoBraneCosmo} we investigate what can be intepreted as emerging brane cosmology.
To this purpose we analyze the associated
 scalar field energy densities, and attempt to distinguish between brane
 localized and bulk energy densities;  this will provide a better
 understanding of the effective dark energy density induced on
 the brane. We also perform a matching
 of the solution of Sec.~\ref{Sec:Solution} to an instantaneous
 RS-like action, to better understand the localization properties
 of the instantaneous bulk and brane cosmological constants.
 We analyze the features of the resulting brane cosmology for early,
 late and intermediate times in Sec.~\ref{Sec:CosmologicalScenarios}, and compare with realistic cosmology.
 Finally, we conclude in Sec.~\ref{Sec:Conclusions}.
\section{General Strategy}
\label{Sec:definiton+solution}
As mentioned in the introduction, previous works on extra
dimensional codimension one domain wall branes (or thick branes) have so far
considered maximally  4-symmetric geometries
($M_4$, $AdS_4$, and $dS_4$). Consequently, the
solutions obtained in this context have limited relevance for cosmology,
which is characterized by a maximally 3-symmetric geometry
described by a  FLRW metric.

The difficulty in obtaining solutions to the cosmological (FLRW)
case stems from the fact that all metric coefficients are $(y,t)$
dependent, where $y$ is the extra dimensional coordinate and $t$
is the time coordinate. The resulting Einstein and Klein-Gordon
equations for the evolving scalar field(s), translate into a system of non linear partial
differential equations, which is extremely difficult to solve even
in the most simplified cases, as we shall see below.

The maximally 4-symmetric DW configurations mentioned above
\cite{DeWolfe,Davidson,GiovanniniM4}, are characterized by a
scalar potential with two degenerate minima and a (static) kink
mode interpolating between the two minima. The nature of the brane
($M_4$, $AdS_4$ or $dS_4$) is studied in the infinitely thin
brane limit~\cite{Davidson}, where the DW is squeezed to its
central point $(y=0$). The  brane type is then determined  from
the brane induced cosmological constant and the bulk cosmological
constant, or more precisely by whether they deviate from the RS
fine tuning relation
\begin{equation}
\kappa(\Lambda_{b})^2/6+\Lambda_5=0
\label{Eq:RSfinetune},\end{equation} where $\Lambda_{b}$
denotes the brane induced cosmological constant coming from the
kinetic and potential terms  and $\Lambda_5$ is the (negative)
bulk cosmological constant, which is identified with the minima of
the scalar potential.

What we wish to emphasize is that in order to study the cosmological
evolution of a 5D time dependent DW configuration, we actually
have to study the evolution of the spatial (3D) geometry along the
$y$ and $t$ directions, simultaneously. This means that in the
most general case, there is no a-priori reason to consider $y=0$
(or any other constant $y$ hypersurface) as the position of the
brane in the extra dimension. In general, we expect there will be
some curve in the $y-t$ plane which is mapped by the DW scalar,
$\phi$, to the center of the DW configuration ($\phi=0$), which is
in turn mapped by $V(\phi)$ to its maximum. Recall that the
existence of a kink mode implies a $Z_2$ symmetry of the potential
which corresponds to two degenerate minima \cite{Davidson}.

\begin{figure}[t]

\begin{center}
\includegraphics[width=6 truecm]{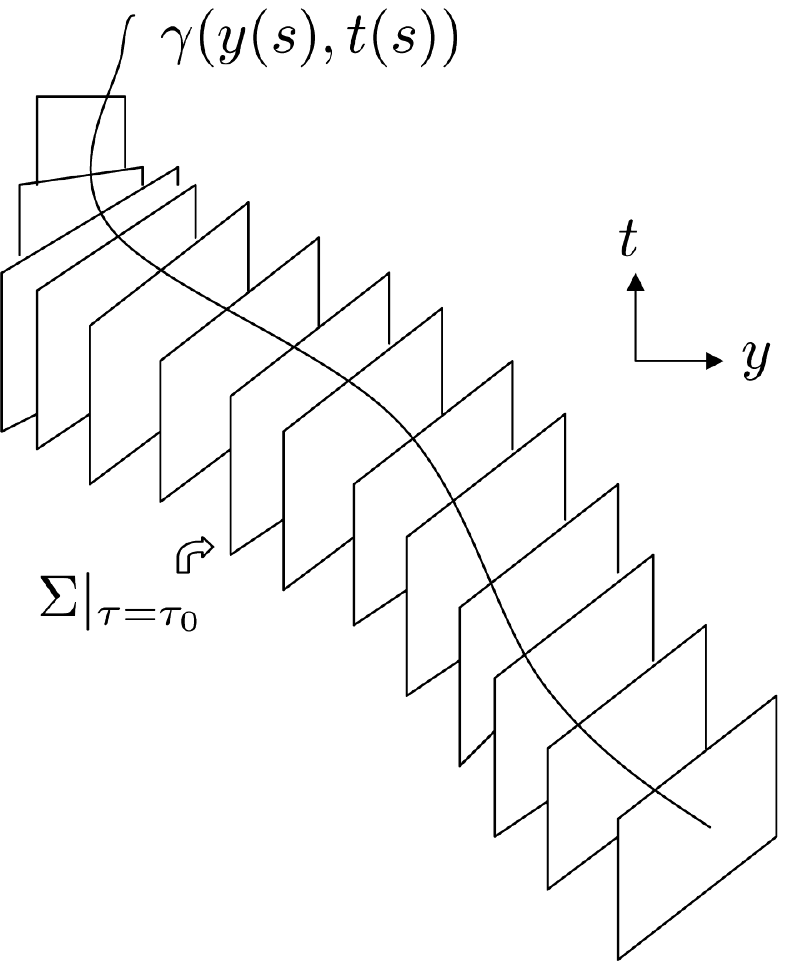}
\caption{{\small Schematic view of the watershed in  $y-t$
coordinates. The curve $\gamma(y(s),t(s))$ represents the points,
which are mapped to the maxima of the scalar potential. The
surfaces $\Sigma|_{\tau=\tau_0}$ denote constant cosmological
time slices, where in the most general case the cosmological time
is identified with the tangent to the curve $\gamma(y(s),t(s))$,
and does not necessarily coincide with the 5D time coordinate
$t$. The cosmological evolution of such a configuration will be
studied by matching the instantaneous bulk and brane
cosmological constants on each of the $\Sigma|_{\tau=\tau_0}$
hypersurfaces.}} \label{Fig:SlinkyScheme}
\end{center}
\end{figure}

The above situation is depicted in Fig.~\ref{Fig:SlinkyScheme},
where the curve $\gamma(y(s),t(s))$ represents the points in the
$y-t$ space, which are mapped to the maxima of the scalar
potential, thereby acquiring the name ``watershed". In the most
general case the tangent to the curve $\gamma(y(s),t(s))$ serves
as the cosmological time, while the normal direction serves
(locally) as the extra spatial coordinate. Namely, at each
constant cosmological time slice, we can study the localization
properties of the kink energy density and distinguish between
brane localized and bulk energy densities. This provides
the effective instantaneous brane cosmological constant,
$\Lambda_b^{eff.}= \kappa\Lambda_b^2/6+\Lambda_5$. In this way, we
actually study the time evolution of a kink induced dark energy
density on a 3-brane located at (and evolving along)
$\gamma(y(s),t(s))$; $\gamma(y(s),t(s))$ is generated by the kink itself. A
fully realistic cosmological evolution will necessitate the
presence of extra sources, for which the most simple example is an
(a-priori) brane localized perfect fluid with equation of state
$p=\omega_f\rho$. These additional sources will correspond to
radiation and matter densities on the brane, while the kink will
contribute the dark energy part. Finally, we wish to  emphasize
that the above prescription is physically plausible, only as long
as the tangent to the curve $\gamma(y(s),t(s))$ is timelike for
any value of the affine parameter $s$.

 The 5D DW
configurations thus generated will be
cosmologically plausible if they can  account simultaneously for a (finite) early time
inflationary period and late time acceleration. This can be achieved by an  induced
brane dark energy density, which is extremely large at very early
times and extremely small at late times.
For a better intuition of their time evolution, we employ a simple
pictorial analogy with a ``slinky" spring:
{\it`` At $t=0$ all the links of a slinky spring, which represents
the DW scalar, are sitting on the maxima of the scalar potential
between two degenerate minima. As time goes by, the links gradually
fall towards the minima on the right and on the left, so that at
 late times, only one link of the slinky is interpolating
between the two piles of links sitting at the two minima"}.

 The
early time accumulation of links at the maxima corresponds to a
huge initial dark energy density, decreasing with time at a yet
unspecified pace. This energy density drives an inflationary
period on the DW brane, which terminates when most of the links of
the slinky hit the minima, or equivalently when the DW kink
configuration becomes very thin. The late time interpolating
link(s) will thus correspond  to a small remnant dark energy
density, which can drive the observed late time acceleration of
our (DW) universe.

We find it important as well as pedagogical to demonstrate, by
means of a simple concrete example, what exactly do we mean by a
slinky evolution. The idea of a slinky evolution has been first
introduced in ref.\cite{OldSlinky}, within the framework of the
so-called geodesic brane gravity. Treating our universe as a
4-dimensional extended object propagating in a 5-dimensional
non-dynamical flat or AdS bulk, its cosmological evolution is then
governed by the Regge-Teitelboim (RT) string-like equations of
motion. In particular, the evolution/nucleation of a de-Sitter
brane was shown to be driven, quite counter intuitively, by a
double-well Higgs potential, namely $V(\phi)=\Lambda +\lambda
(\phi^2-v^2)^2$, rather than by a plain cosmological constant.
Using the static radially symmetric representation of the
de-Sitter metric
\begin{equation}
    ds^{2} = -\left(1-\textstyle{\frac{1}{3}}
    \Lambda R^{2}\right)dt^{2} +
    \frac{dR^{2}}{\left(1-\frac{1}{3}\Lambda R^{2}\right)} +
    R^{2}d\Omega^{2} ~,
\label{Eq:DesitterRadial}\end{equation} and reflecting some novel
seesaw interplay between the cosmological energy density and its
effective RT companion, the corresponding time dependent evolution
of the associated scalar field was derived to be
\begin{equation}
 \phi(t,R) = \frac
    {v\sqrt{1-\frac{1}{3}\Lambda R^{2}}
    \sinh\sqrt{\frac{\Lambda}{3}}t}
    {\sqrt{1+\left(1-\frac{1}{3}\Lambda R^{2}\right)
    \sinh^{2}\sqrt{\frac{\Lambda}{3}}t}} ~.\label{Eq:oldslinky}
\end{equation}
At $t=0$, all points in space share a common $\phi(0,R)=0$. At $t>
0$, however, it is exclusively on the event horizon
$R=\sqrt{\Lambda/3}$ where the scalar field, experiencing an
infinite gravitational red-shift, gets frozen in its unbroken
phase. As $t\rightarrow\infty$, and for every point $R$ in space
excluding $R=\sqrt{\Lambda/3}$, the scalar field smoothly connects
$\phi(-\infty,R)\rightarrow -v$ with $\phi(\infty,R)\rightarrow
v$. This concludes the demonstration of the slinky evolution. In
the present paper, we adopt the general idea of the slinky
evolution in an attempt to account for a slinky creation of a FRW
brane.

Given these requisites,  the simplest slinky solution we can think of are scalar profiles,
nearly flat along the extra dimension at early times and evolving to
step functions at late times.  The location of the jump or
equivalently, the center of the DW configuration, will correspond
to the position of our brane universe in the extra dimension.
Thus, the thin brane limit of slinky configurations is achieved
dynamically at late times, instead of being a  mathematical
limit of the parameters entering a static DW solution, as in
the maximally 4-symmetric cases considered in \cite{Davidson}.

A feature that cosmologically
plausible slinky configurations should also  possess is the ability to
generate a ``brane in time" in addition to a ``brane in the extra
dimension". Namely, in order for such configurations to account
for the creation of the evolving brane universe described above,
the hypersurface $\Sigma|_{\tau=0}$, corresponding to zero
cosmological time, should be distinguished from  all  other
constant $\tau$ surfaces, and identified with  the thin brane limit in the
(cosmological) time direction.

A configuration  that possesses  the above highly nontrivial
property will naturally correspond to the Hartle-Hawking no
boundary proposal \cite{Hartle-Hawking}.

\section{ Setup and the simplest slinky configuration}\label{Sec:Solution}

The only recorded attempt at finding time dependent DW solutions
is in \cite{Giovannini}, where the simplest ansatz for the
geometry is employed, with  a single warp factor depending on time
and the extra dimensional coordinate.  As a first step towards
obtaining DW slinky configurations, we seek for a solution with
the desired properties and for the same metric ansatz.

We start by defining the setup. The 5D action is given by:
\begin{equation} S=\int
d^5x\sqrt{G}\left[-\frac{R}{2\kappa}+\frac{1}{2}G^{AB}\partial_A\phi\partial_B\phi-V(\phi)\right]\,
,\label{action}\end{equation} with metric ansatz
\begin{equation}
ds^2=G_{AB}dx^Adx^B=a^2(\omega,\tau)\left[d\tau^2-d\vec{x}^2\right]-b^2(\omega,\tau)d\omega^2
\label{metric},\end{equation} where $\kappa=8\pi G_5=8\pi/M_5^3$
and $G_5$ is the 5D gravitational constant. The coordinates
$\omega$ and $\tau$ are conformal coordinates, which are related
to the more commonly used proper distance and proper
(cosmological) time coordinates by the transformations
$b(\omega,\tau)d\omega=dy$ and
$a(\omega,\tau)d\tau=dt$.\footnote{The $\omega$ coordinate can
still be thought of as a proper distance coordinate for the metric
ansatz of Eq.~(\ref{metric}), but we choose to treat it as
conformal due to symmetry arguments which are made clear below.}
The Einstein equations for the action in Eq.~(\ref{action}) can be
written as follows:
\begin{equation}
R_{AB}=\kappa\left[\partial_A\phi\partial_B\phi-\frac{2}{3}G_{AB}V\right]
\label{EE}\end{equation}
We explicitly write the components of the Einstein equations in a form analogous to
\cite{Giovannini}. For the $(00)$ component we get:
\begin{equation}
\frac{a^2}{b^2}\left[h^{\prime}+4h^2-hf\right]-\left[\dot{F}+3\dot{H}+F(F-H)\right]=
\kappa\left[\dot{\phi}^2-\frac{2}{3}Va^2\right],
\label{EE00}\end{equation} where $()^{\prime}$ denotes
differentiation with respect to $\omega$ and $\dot{()}$ denotes
differentiation with respect to $\tau$. In addition  we have
defined $H(\omega,\tau)=\dot{a}/a$, $h(\omega,\tau)=a^{\prime}/a$,
$F(\omega,\tau)=\dot{b}/b$ and $f(\omega,\tau)=b^{\prime}/b$. For
the $(ij)$ ($i,j=1,2,3$) component we obtain (we omit
$\delta_{ij}$):
\begin{equation}
-\frac{a^2}{b^2}\left[h^\prime+4h^2-hf\right]+\left[\dot{H}+2H^2+HF\right]=\frac{2}{3}\kappa
Va^2.\label{EEij}\end{equation}
The $(55)$ component yields:
\begin{equation}
\frac{b^2}{a^2}\left[\dot{F}+F^2+2HF\right]-4h^{\prime}+4h\left(f-h\right)=\kappa\left[\phi^{\prime
2}+\frac{2}{3}Vb^2\right]. \label{EE55}\end{equation}
Finally, the $(05)$ component provides a momentum
constraint (on $\phi$) of the following form:
\begin{equation}
\dot{h}-4H^{\prime}+3Fh=\kappa\dot{\phi}\phi^{\prime}\, .\label{EE05}\end{equation}
The $(00)$ and $(ij)$ equations can be combined to yield:
\begin{equation}
\dot{\phi}^2=\frac{1}{\kappa}\left[2H^2+2HF-F^2-\dot{F}-2\dot{H}\right].\label{EEphidot}\end{equation}
Similarly, if we combine the $(ij)$ and $(55)$ components we obtain
\begin{equation}
\phi^{\prime
2}=\frac{3}{\kappa}\left(hf-h^\prime\right)+\frac{1}{\kappa}\frac{b^2}{a^2}\left[\left(\dot{F}-\dot{H}\right)
+\left(F+2H\right)\left(F-H\right)\right].\label{EEphiprime}\end{equation}
Turning to the Klein-Gordon equation
($G^{AB}\nabla_A\nabla_B\phi+\partial V/\partial\phi=0$) and using
the metric in Eq.~(\ref{metric}) we get:
\begin{equation}\ddot{\phi}-\frac{a^2}{b^2}\phi^{\prime\prime}+\left(2H+F\right)
\dot{\phi}-\frac{a^2}{b^2}(4h-f)\phi^\prime+\frac{\partial
V}{\partial\phi}a^2=0\, . \label{KG}\end{equation}
Having stated the relevant equations we look for the most simple
realization of the slinky type configuration introduced in Sec. 2.
The first type of solutions we are going to look for correspond to
the simplest watershed possible, namely, a line of constant
$\omega$ that we choose to be $\omega=0$ without loss of
generality.

\noindent We first recall the time shifted solutions of
\cite{Giovannini}. Generalizing the (static) maximally 4-symmetric
solution of \cite{GiovanniniM4} to the time dependent case by the
simple coordinate redefinition,
$\omega\to\omega+\tau$\footnote{This coordinate redefinition is
equivalent to a $\pi/4$ rotation of the static solution of
\cite{GiovanniniM4} in the $\omega-\tau$ plane.}, the author of
\cite{Giovannini} was able to show that the following geometry,
kink profile and scalar potential solve the Einstein and
Klein-Gordon equations (Eqs.(\ref{EE00})\,--\,(\ref{EE05}) and
(\ref{KG})):
\begin{eqnarray}
b(\omega,\tau)&=&\epsilon
a(\omega,\tau)=\epsilon\frac{1}{\sqrt{1+\lambda^2(\omega+\tau)^2}}\,,\quad
\phi(\omega,\tau)=\sqrt{\frac{3}{\kappa}}\arctan[\lambda(\omega+\tau)]\,,\nonumber\\
V(\Phi)&=&\frac{3\lambda^2}{2\kappa}\left(\frac{1-\epsilon^2}{\epsilon^2}\right)(1-5\sin^2\Phi),
\label{Eq:GiovanniniOriginal}
\end{eqnarray}
where $\Phi\equiv\sqrt{\kappa/3}\,\phi$. The cosmology associated
with the above solution was studied in \cite{Giovannini} by
inspecting  $H(0,\tau)$: the latter exhibits a bouncing behavior around
$\tau=0$, yet of sign opposite to the one of a realistic cosmological
evolution. In addition, it was shown that in the single field case
a factorizable dependence of the warp factors on $\omega$ and
$\tau$ implies trivial cosmology, namely $H(0,\tau)=0$. The
inclusion of an additional scalar, $\chi$, allows for a non
trivial cosmological evolution driven by its kinetic term. In this
setting, $\chi$ is a purely $\tau$ dependent free field, while the
DW field, $\phi$, is purely $\omega$ dependent \cite{Giovannini}.

Two aspects of the single field configuration in
Eq.~(\ref{Eq:GiovanniniOriginal}) are relevant to our purpose. The
first is that the kink configuration is in this case always
centered around $\omega+\tau=0$, thus $\omega=0$ will actually
correspond to the location of the DW brane only at $\tau=0$. More
in general, the concept of watershed as introduced in
Sec.~\ref{Sec:definiton+solution} is not applicable, as no
evolution of the shape or width of the DW configuration is
experienced along the curve $\omega+\tau=0$.\newline\noindent
Secondly, the solution in Eq.~(\ref{Eq:GiovanniniOriginal}), as
well as the static solution of \cite{GiovanniniM4} with kink
profile $\phi(\omega)=\sqrt{3/\kappa}\,arctan[\lambda\omega]$, are
written in conformal coordinates and it is instructive to study
its properties in proper distance coordinates. Integrating the
relation $a(\omega,\tau)d\omega=dy$ and rescaling $y\to \epsilon
y$, we rewrite the metric for the solution of
Eq.~(\ref{Eq:GiovanniniOriginal}) in $(y,\tau)$ coordinates 
\begin{equation}
ds^2=-b^2(\omega,\tau)d\omega^2+a^2(\omega,\tau)\left[d\tau^2-d\vec{x}^2\right]=-dy^2+
{\rm sech}^2 \left(\lambda y\right)\left[d\tau^2-d\vec{x}^2\right]
\label{TimeShiftedMetric}.
\end{equation} 
The above metric describes a static 5D warped spacetime which is
asymptotically $AdS_5$. If we would have transformed to a proper
time coordinate, $t$ instead of $y$, by requiring
$a(\omega,\tau)d\tau=dt$, we would have instead gotten a metric
warped in time (with warp factor $a(y,\tau)=sech[\lambda\,t])$,
for which the cosmological interpretation is that of an apparent
reversed bouncing behavior as discussed  above and in
\cite{Giovannini}.

In order to obtain more realistic solutions in the spirit of the
slinky framework, we find it instructive to consider kink profiles
and geometries which exhibit a more generic non-factorizable time
dependence. As a first attempt, we propose a time ``proportional"
kink solution, with a non factorizable dependence on $\omega$ and
$\tau$. A simple example of such a kink configuration is:
\begin{equation}
\phi\propto \arctan[\lambda \omega\tau].
\label{kinkansatz}\end{equation}
It satisfies $\phi(\omega,\tau=0)=0$, which we interpret as the
initial state of a slinky DW configuration, where all  the links
of the slinky ($\phi$ values along $\omega$ at a constant $\tau$
slice) are sitting on the maxima of the yet to be determined
potential.  The  time evolution of the above configuration
dictates that as time goes by more and more links of the slinky
will fall towards the minima of the potential as $\phi$ approaches
its asymptotic value. Although the potential and warp factors are
yet unspecified, it is clear that a kink solution will require a
$Z_2$ symmetry of the potential, which corresponds to two
degenerate minima symmetric around $\phi=0$. For the kink
configuration of Eq.~(\ref{kinkansatz}) the $Z_2$ symmetry
translates into reflection symmetry around $\omega=0$, under which
the kink has to be odd. In addition, since we are interested in
the localization of gravity to the dynamically generated domain
wall brane universe,  we should consistently seek for warp factors
that are peaked at $\omega=0$ and are $Z_2$ even.

We now consider the  time ``proportional" generalization of
the maximally 4-symmetric warp factors in the
spirit of the time shifted solutions of \cite{Giovannini}:
\begin{equation}
a(\omega,\tau)=\frac{1}{\sqrt{1+\lambda^2\omega^2\tau^2}} \qquad
b(\omega,\tau)=\epsilon a(\omega,\tau)\, .\label{warpansatz}
\end{equation}

Substituting the above warp factors in Eq.~(\ref{EEphidot}) and
integrating, we obtain a time proportional solution for $\phi$:
\begin{equation}
\phi=\sqrt{\frac{3}{\kappa}}\arctan\left[\lambda
\omega\tau\right]+f_1(\omega) \label{kinksolution},\end{equation}
with integration constant $f_1(\omega)$.
Substituting again Eq.~(\ref{warpansatz}) in
Eq.~(\ref{EEphiprime}), we similarly obtain:
\begin{equation}\phi=\sqrt{\frac{3}{\kappa}}\arctan\left[\lambda
\omega\tau\right]+f_2(\tau),\end{equation}
with integration constant $f_2(\tau)$. Substituting the solution
of Eq.~(\ref{EEphiprime}) in Eq.~(\ref{EEphidot}) and vice versa,
we realize that $f_1(y)=f_2(\tau)=const$. Setting
$f_1(\omega)=f_2(\tau)=0$, we substitute the above solutions for
$\phi$, $a$ and $b$ in Eq.~(\ref{EEij}),  solve for the potential
$V(\phi)$ and obtain:\begin{eqnarray}
V(\Phi)&=&\frac{3\lambda^2(\tau^2-\epsilon^2\omega^2)}{2\kappa\epsilon^2}\left
(1-5\sin^2\Phi\right
)=\frac{3(\lambda^2\tau^2-\epsilon^2\tan^2\Phi/\tau^2)}{2\kappa\epsilon^2}\left
(1-5\sin^2\Phi\right
)\nonumber\\&=&\frac{3(\tan^2\Phi/\omega^2-\lambda^2\epsilon^2\omega^2)}{2\kappa\epsilon^2}\left
(1-5\sin^2\Phi\right )\label{potential},
\end{eqnarray}
where $\Phi=\sqrt{\kappa/3}\phi$. The above solution for $V(\Phi)$
is consistent with the rest of the Einstein equations in which $V$
appears and the Klein-Gordon equation.
However, it is affected by what we are used to consider a highly
problematic feature, that is it contains an explicit coordinate
dependence\footnote{Notice that the coordinate dependence  renders
the potential and thus the whole solution to be non
$\omega\leftrightarrow\tau$ symmetric, a feature that might be
welcome in distinguishing time evolution from space dynamics.},
thus violating 5D general covariance. As a result, the momentum
constraint given by the $(05)$ component of the Einstein equation
Eq.~(\ref{EE05}) is not satisfied, as it corresponds to 5D
conservation of energy and momentum, which is itself a result of
5D covariance.

Specifically, for the kink in Eq.~(\ref{kinksolution}) and
the warp factors in Eq.~(\ref{warpansatz}), the two sides of Eq.~(\ref{EE05}) for the
$(05)$ component take the following form:
\begin{equation}
\dot{h}-4H^{\prime}+3Fh=\frac{3\lambda^2\omega\tau}{(1+\lambda^2\omega^2\tau^2)^2}
+\frac{3\lambda^2\omega\tau}{(1+\lambda^2\omega^2\tau^2)} \quad
\neq \quad
\kappa\dot{\phi}\phi^{\prime}=\frac{3\lambda^2\omega\tau}{(1+\lambda^2\omega^2\tau^2)^2}\,.
\label{05Violation}\end{equation}
Importantly, notice
that as $\omega\to 0,\infty$ and/or $\tau\to 0,\infty$ the
violation of the (05) equation vanishes, precisely in those
regions where our configuration mimics a brane in the extra
dimension or in time. In addition, the violation of the $(05)$
equation is odd in $\omega$ and $\tau$ and will vanish if we
integrate one (or both) of them out. It is clear that as long as
one considers a potential $V(\phi)$ with an explicit coordinate
dependence, local properties cannot be inferred from 5D covariance
away from the asymptotic regions.

The apparent physical meaning of the violation of the above
equation is that the energy flux along the 5th dimension (or
equivalently the 5 dimensional momentum density) encoded in
$T_{05}^\phi$ is not enough to account for the analogous flux
associated with $G_{05}$; namely, the kink solution in
Eq.~(\ref{kinksolution}) does not inject enough energy in the 5th
dimension in order to support the geometry described by the warp
factors in Eq.~(\ref{warpansatz}). This may suggest that a
possible cure may be provided by the addition of a scalar density
source or dynamical scalar field to the minimal framework
discussed here. Another possibility worth to explore is the
presence of non-canonical kinetic energy terms for the scalar
field(s). We stress that, while the physics at hand suggests that
a fully realistic brane cosmology should possibly result from a
non-factorizable dependence upon time and the extra spatial
dimension, no solution to the complete problem has been found
until now - with the exception of the time-shifted solution in
\cite{Giovannini} which can however be interpreted as a rotation
of the static solution and it does not provide a realistic
cosmology; the problematic features that we are encountering,
related to the way the temporal and extra spatial coordinates are
entangled, may suggest the way towards an improved description.

Despite the outlined problems, it is an instructive exercise to
further investigate the time evolution generated by this first
attempt, and understand to what degree it provides a reasonable
cosmology.  In particular, we try to better understand the
dynamics of the slinky links (portions of the slinky configuration
itself) on the terrain of the scalar potential. 

\begin{figure}[]
\subfigure[Early times $(\tau_0\to 0)$]{
\includegraphics[scale = .6]{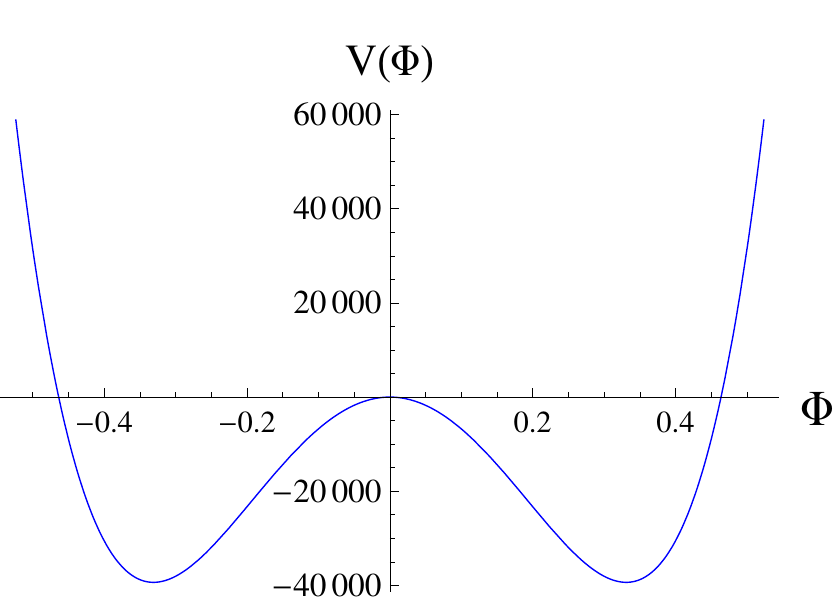}
\label{pot1} } \subfigure[Intermediate
times\,$(\tau_0\sim\mathcal{O}(1))$]{
\includegraphics[scale = .6]{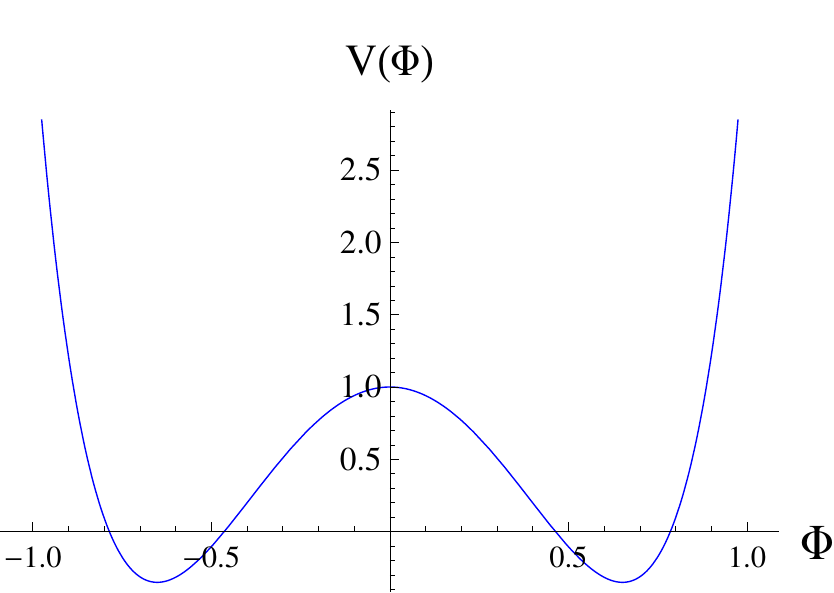}
\label{kk} } \subfigure[Late times $(\tau_0 >>1)$]{
\includegraphics[scale = .6]{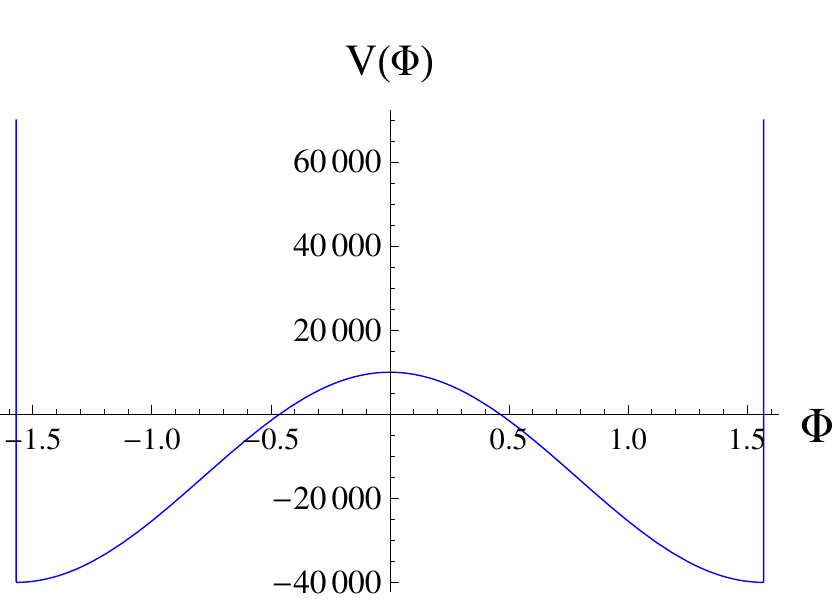}
\label{k1} }  \caption{The scalar potential on the constant $\tau$
slices, $\Sigma|_{\tau=\tau_0}$
(Eq.~(\ref{PotentialConstantSlices})), for early, intermediate and
late times. We set $\kappa=\lambda=1$ and $\epsilon=0.9$. Notice
the increased depth of the two degenerate minima in early and late
times and the rise of the maxima in late times.}
\label{Fig:PotentialSlices}
\end{figure}

To this aim, it is useful to inspect the potential on constant $\tau$-slices, the
$\Sigma|_{\tau=\tau_0}$ hypersurfaces, where we can
rewrite $V(\Phi)$ of Eq.~(\ref{potential}) covariantly as follows:
\begin{equation}
V(\Phi)=\frac{3\lambda^2}{2\kappa\epsilon^2}\left(\frac{\tilde{\lambda}^2}{\lambda^2}
-\frac{\epsilon^2}{\tilde{\lambda}^2}(\tan^2\Phi)\right)(1-5\sin^2\Phi),
\label{PotentialConstantSlices}
\end{equation}
where $\Phi\equiv\sqrt{\kappa/3}\phi$ and $\tilde{\lambda}=\lambda\tau_0$. The above potential is still of a
double well form for small values of $\tilde{\lambda}$, yet the
location of the minima is different on each constant $\tau$ slice
and overlaps the asymptotic value $\Phi=\pm\pi/2$ only in the
$\tau\to\infty$ limit. In this limit, $V(\Phi)$ evolves to a
``chopped double-well" shape, where the edges become the location of
both the minima and the infinitely high walls of $V(\Phi)$,
corresponding to the asymptotic values $\Phi=\pm\pi/2$. We depict
the form of the potential for early, intermediate and late times
in Fig. \ref{Fig:PotentialSlices}. The location of the
majority of the links will always correspond to the potential in
the vicinity of the asymptotic values, $\Phi=\pm\pi/2$, except for
the $\tau=0$ hypersurface, where all the links sit
on the local maxima of $V(\Phi)$ at $\Phi=0$.

We conclude that most of the links of the slinky  experience a
fall from the two walls  of $V(\Phi)$ (around $\Phi=\pm\pi/2$)
towards the minima, the latter approaching the $\Phi=\pm\pi/2$ values as
$\tau\to\infty$. Hence, it is only at late times that we can
interpret the scalar profile in Eq.~(\ref{kinksolution}) as a true
kink interpolating between the two (cusped) degenerate minima
of $V(\Phi)$, as speculated in Sec.~\ref{Sec:definiton+solution}.


\section{From Domain Wall to Brane cosmology}
\label{Sec:DWtoBraneCosmo}

In this section we elaborate on what it is exactly that we are going to
interpret as brane cosmology in the above setup.

Observing the time evolution of $\phi(\omega,\tau)$, we realize
that when $\tau\to\infty$ we are naturally in a thin brane limit,
where the brane is  located at $\omega=0$. Inspired by the thin
brane limit of maximally 4-symmetric domain wall configurations
with non vanishing brane cosmological constant, $\Lambda_b$
\cite{Davidson}, we recall the way in which this cosmological
constant is matched. To be more specific the brane cosmological
constant is due to the energy density of the bulk soliton when it
is squeezed to the very same brane it generates at $\omega=0$,
once the thin limit is taken.

\subsection{Obtaining the brane cosmological constant in the thick
RS case}\label{Sec:LambdaM4}

As an instructive exercise, we review the results of
\cite{Davidson} for a maximally 4-symmetric domain wall
configuration which admits the RS model as its thin limit. The
metric ansatz is:
\begin{equation}
ds^2=-dy^2+e^{2f(y)}\left[dt^2-d\vec{x}^2\right].
\label{M4metric}\end{equation}
 Considering again a minimally coupled scalar field $\varphi$ in the
 geometry described by the above equation, one writes the Einstein
 and Klein-Gordon equations, which are significantly simpler than
 the time dependent case and obtains the following solution:
 \begin{equation}
 \varphi=A\arctan[\tanh(\beta y/2)] \qquad
 V(\varphi)=A^2\beta^2/8-(A^2\beta^2/8)(1+\kappa A^2/3)\sin^2(2\varphi/A)
 \label{M4kinkPot},\end{equation}
 where $A$ and $\beta$ are constants. The geometry that
 corresponds to the above kink configuration and potential is
dictated
 by $e^{f(y)}$, where:
 \begin{equation}
 e^{f(y)}=D[\cosh(\beta y)]^{-\kappa A^2/12}
 \label{M4Geom}.\end{equation}

 The thin limit of the above configuration is a  delicate one
 in which
$\beta\to\infty$, $A\to 0$ and $A^2\beta$ is fixed. In this limit
$f\to -\kappa A^2\beta|y|/12$, such that $f^\prime\to-(\kappa
A^2\beta/12)(\theta(y)-\theta(-y))$ and
$f^{\prime\prime}\to-(\kappa A^2\beta/6)\delta(y)$. The
resulting bulk geometry is that of a slice of $AdS_5$
divided by the brane (infinitely thin domain wall centered at
$y=0$) into two disconnected regions, across which there is a
jump of the extrinsic curvature. On identifying the degenerate
minima of the potential in Eq.~(\ref{M4kinkPot}) with the 5D
cosmological constant associated to the thin-limit $AdS_5$ space,
one obtains:
\begin{equation} V(\pm \pi A/4)=\Lambda_5=-\kappa A^4\beta^2/24
\quad \Rightarrow \quad f^{\prime\prime}\to
-2(-\Lambda_5\kappa/6)^{1/2}\delta(y)
\label{thinLambda}.\end{equation}
The presence of the delta function in $f^{\prime\prime}$ combined
with the (00) component of the Einstein equations for the metric
in Eq.~(\ref{M4metric}) \cite{Davidson}, entail that both
$\varphi^{\prime 2}$ and $V(\varphi)$ must  contain the same
delta function. Consequently, in the thin limit a term
$e^{2f}[\varphi^{\prime 2}/2+V(\varphi)]\to
A^2\beta/2\delta(y)=\Lambda_b\delta(y)$ is generated in $T_{00}$.
This term  corresponds to a brane cosmological constant, which
automatically satisfies the RS fine tuning relation
$\Lambda_5+\kappa\Lambda_b^2/6=0$. Namely, what had to be put by
hand in the RS model is an output of the above dynamics.

An alternative way to derive $\Lambda_b$ is by requiring the 5D
action $\int dy\sqrt{-g}\mathcal{L}=-\int dy
e^{4f}[\varphi^{\prime 2}/2$\newline$+V(\varphi)]=(3/\kappa)\int
dy e^{4f}(f^{\prime\prime}+2f^{\prime 2})$ to generate the RS
action $\int dy e^{4f}(-\Lambda_5-\Lambda_b\delta(y))$ in the thin
limit. Using this prescription it is clear that $\Lambda_5$ comes
only from the $2f^{\prime 2}$ term and $\Lambda_b$ comes only from
the $f^{\prime\prime}$ term. The RS fine tuning relation is
obviously satisfied by these results.
\subsection{Obtaining the brane cosmological constant for
slinky configurations}\label{Sec:LambdaSlinky}
In our case we will assume that the time evolution of the kink
configuration of Eq.~(\ref{kinksolution}), also
controlled by the dimensionless parameter $\lambda$, is slow
enough that a thin brane limit
can be taken at every constant time slice,  similarly to the maximally 4-symmetric case of
\cite{Davidson}. Then, by matching to an instantaneous RS-like
action or inspecting the behavior of $T_{00}$, we determine
the  bulk ($\Lambda_5(\tau)$) and brane induced
($\Lambda_b(\tau)$) cosmological constants at each slice of
constant $\tau$. The brane (bulk) cosmological constant is an
energy density which is constant in the instantaneous 3(4) space
dimensions and satisfies $p=-\rho$.

 To further elaborate along the lines of  the above procedure, we first write the metric components along
a constant $\tau$ slice, at say $\tau=\tau_0$:
\begin{equation}
\left.ds^2\right|_{\tau=\tau_0}=g^{\Sigma_0}_{ij}dx^idx^j=-a^2(\omega,\tau=\tau_0)d\vec{x}^2-\epsilon^2
a^2(\omega,\tau=\tau_0)d\omega^2=\frac{-1}{{1+\tilde{\lambda}^2\omega^2}}[d\vec{x}^2+\epsilon^2d\omega^2]
\label{slicemetric},\end{equation}
where $\tilde{\lambda}=\tau_0\lambda$ and spatial indices
$i,j=x,y,z,\omega$. We note in passing that the metric ansatz we
chose in Eq.~(\ref{metric}) is conformally flat, but in contrast
with the maximally 4-symmetric cases in
\cite{DeWolfe,Gremm:Dw:1999pj,Gremm:dS4:2000dj,Davidson,GiovanniniM4},
it does not admit an asymptotic $AdS_5$ limit (with a 5D
cosmological constant corresponding to the value of the minima of
the potential), due to the non-removable coordinate dependence in
$V(\phi)$. Moreover, although we have written the time and extra
dimension as $\tau$ and $\omega$, respectively, we recall that the
$\tau-\omega$ symmetric ansatz for the warp factors in
Eq.~(\ref{warpansatz}), is inspired by the solutions to the
maximally 4-symmetric case with a conformal extra dimensional
coordinate \cite{Giovannini}. Thus, in order to inspect the
cosmological time, $t$, and proper distance coordinate, $y$, we
should in principle rewrite the solution using the relations
$dy=b(\omega,\tau)d\omega$ and $dt=a(\omega,\tau)d\tau$
\footnote{We have not specified whether $\tau$ is indeed a
conformal coordinate, but merely stated that the ansatz for $a$
and $b$ is inspired by a solution to the maximally 4-symmetric
case, where the $\omega$  coordinate is conformal. This is also
true for $\tau$ in the ``time shifted" solutions of
\cite{Giovannini}.}. Provided no singularity occurs in the
coordinate transformation, we can work with $(\omega,\tau)$
coordinates and later transform to $(y,t)$ coordinates without
loss of generality.

Considering the instantaneous geometry on each constant time
slice described by the metric in Eq.~(\ref{slicemetric}), we
obtain the bulk energy density and match it to a RS-like setup as
follows. From the action in Eq.~(\ref{action}), or equivalently from
the (00) component of the energy momentum tensor
$T_{00}=(1/2)\dot{\phi}^2+(1/2\epsilon^2)\phi^{\prime
2}+a^2V(\phi)$ (where $T_{MN}=(2/\sqrt{G})\delta\mathcal{L}/\delta
G^{MN}$ and the Einstein Equations take the form
$R_{MN}-(1/2)G_{MN}R=\kappa T_{MN}$), we realize that we have
three separate contributions to the energy density which we
label as: $\Omega_{\phi^{\prime 2}}$, $\Omega_{\dot{\phi}^2}$ and
$\Omega_{V(\phi)}$, where $T_{00}=\Omega_{\phi^{\prime
2}}+\Omega_{\dot{\phi}^2}+\Omega_{V(\phi)}$ and
\begin{equation}\label{DensitiesDefprime}
\Omega_{\phi^{\prime 2}}=\frac{1}{2\epsilon^2}\phi^{\prime
2}=\frac{3\tau^2\lambda^2}{2\kappa\epsilon^2(1+\lambda^2\omega^2\tau^2)^{2}}=
\frac{3\tilde{\lambda}^2}{2\kappa\epsilon^2(1+\tilde{\lambda}^2\omega^2)^{2}},
\end{equation}
\begin{equation}\label{DensitiesDefdot}
\Omega_{\dot{\phi}^2}=\frac{1}{2}\dot{\phi}^2=
\frac{3\omega^2\lambda^2}{2\kappa(1+\lambda^2\omega^2\tau^2)^{2}}=
\frac{3\omega^2\lambda^2}{2\kappa(1+\tilde{\lambda}^2\omega^2)^{2}},
\end{equation}
\begin{equation}\label{DensitiesDefPotential}
\Omega_{V(\phi)}=a^2V(\phi)=\frac{3 \lambda ^2
\left(\tau^2-\omega^2 \epsilon ^2\right) \left(1-4 \tau^2 \omega^2
\lambda ^2\right)}{2 k \epsilon^2  \left( 1+ \tau^2 \omega^2 \lambda
   ^2\right)^{2}}=\frac{3
\left(\tilde{\lambda}^2-4\tilde{\lambda}^4\omega^2-\lambda^2\omega^2
\epsilon^2+4\tilde{\lambda}^2\lambda^2\omega^4\epsilon^2\right)}{2
k \epsilon^2 \left( 1+\tilde{\lambda}^2\omega^2\right)^{2}}.
\end{equation}
In the above equations the second equality is written on
$\Sigma|_{\tau=\tau_0}$, on which we are going to characterize
each of the above contributions, term by term, to determine the
localization properties of the associated energy densities and
consequently what should be interpreted as $\Lambda_b(\tau_0)$.

To implement the procedure described in the above paragraph, we
perform a transformation $\omega\to y$ to a proper distance extra
dimensional coordinate defined on each constant time slice,
$\Sigma|_{\tau=\tau_0}$. Using Eq.~(\ref{slicemetric}) we obtain:
\begin{equation}
a(\omega, {\tau=\tau_0}) d\omega=\frac{1}{\sqrt{1+\tilde{\lambda}^2\omega^2}}d\omega=dy
\quad \Rightarrow \quad y= {\rm arcsinh}
(\tilde{\lambda}\omega)/\tilde{\lambda} \quad \Rightarrow \quad
\omega=\sinh (\tilde{\lambda}y)/\tilde{\lambda}
\label{conformalTrans}.\end{equation}
While the scaling of each term in the energy densities in
Eqs.~(\ref{DensitiesDefprime})-(\ref{DensitiesDefPotential}) with
the time coordinate $\tau$ is determined by the powers of
$\tilde{\lambda}$, their $\omega$ dependence remains explicit and
is translated into the dependence on $y$ according to
Eq.~(\ref{conformalTrans}). The thin limit on each constant $\tau$
 slice (when $\tau_0$ is simply a finite constant), is taken in
principle by letting $\lambda\to\infty$. However, this limit is not well
defined since the $\lambda$ dependence of all contributions is
such that most of them will naively vanish in this limit. We can
in principle overcome this problem by modifying $\phi$,
$V(\phi)$ and $a(\omega,\tau=\tau_0)$ to contain an additional
pre-factor $A$ such that the thin limit is taken by keeping
$A^2\lambda$ fixed while letting $A\to 0$ and $\lambda\to\infty$
as in \cite{Davidson}. However, such  a modification of the warp
factor, $a(\omega,\tau=\tau_0)$, by virtue of
Eqs.~(\ref{EEphidot}) and (\ref{EEphiprime}), will result in a non
solitonic profile which blows towards infinity and will therefore
fails to generate the thick brane (kink) we are looking for in
the first place.

 On the other hand, by
simply observing the $\omega$ dependence of the same contributions
in Eqs.~(\ref{DensitiesDefprime})-(\ref{DensitiesDefPotential}),
we see that they are all (except for the last term in
Eq.~(\ref{DensitiesDefPotential})) peaked at (or around)
$\omega=0(y=0)$. In particular, the common denominator for all
contributions, $\tilde{D}( \tilde{\lambda},\omega)\equiv
1/(\sqrt{1+\tilde{\lambda}^2\omega^2})^4$ translates into $
[sech(\tilde{\lambda}y)]^4$, which is strongly peaked at $y=0$
when $\tilde{\lambda}\to\infty$. Moreover, the $\omega$ dependence
of $\Omega_{\phi^{\prime 2}}$, $\Omega_{\dot{\phi}^2}$ and
$\Omega_{V(\phi)}$  is characterized by three distinct
behaviours, $\tilde{D}$, $\omega^2\tilde{D}$ and
$\omega^4\tilde{D}$, each of which also enters with different
powers of $\tilde{\lambda}$. Since $\omega=
\sinh(\tilde{\lambda}y)/\tilde{\lambda}$, it is clear that the
$\omega^2\tilde{D}$  terms will also be peaked around $\omega=0$.
 So, a priori it seems like all of the
energy densities in
Eqs.~(\ref{DensitiesDefprime})-(\ref{DensitiesDefPotential})
should be interpreted as contributions to the  brane cosmological
constant and do not correspond to any bulk cosmological
constant, except  for the fourth term in $\Omega_V$. In order to
determine the brane induced cosmological constant we will
integrate the densities of
Eqs.~(\ref{DensitiesDefprime})-(\ref{DensitiesDefPotential}) over
$\omega$ on $\Sigma|_{\tau=\tau_0}$, to get a result
which we are going to interpret as the instantaneous cosmological
constant on the 3-brane located at $\omega=0$, at $\tau=\tau_0$.
Labelling the four terms in $\Omega_V$ of
Eq.~(\ref{DensitiesDefPotential}) as $\Omega_V^{1,2,3,4}$, we
realize that $\Omega_{\phi^{\prime 2}}=\Omega_{V}^{1}$ and
$\Omega_{\dot{\phi}^2}=-\lambda^2\epsilon^2\Omega_{V}^2/(4\tilde{\lambda}^4)=-\Omega_V^3$.

The limits of integration $\omega\to\pm\infty$ translate into
$y\to\pm\infty$, as it can be inferred from  Eq.~(\ref{conformalTrans}).
Starting from $\Omega_{\phi^{\prime 2}}$ and recalling that
$d\omega=\cosh(\tilde{\lambda}y)dy$ we get:\begin{equation} \int
d\omega\sqrt{G}\Omega_{\phi^{\prime 2}}=\int dy
\frac{3\tilde{\lambda}^2}{2\kappa\epsilon (\cosh^8(\tilde{\lambda}y))}=\frac{3\tilde{\lambda}^2}{2\kappa\epsilon}
\left(\left.\frac{16
\tanh(\tilde{\lambda}y)}{35\tilde{\lambda}}\right|^{y_+\to+\infty}_{y_-\to-\infty}
+ \ldots\right)\approx \frac{48\tilde{\lambda}}{35\kappa\epsilon}
,\label{phiprimeintegral}
\end{equation}
where the $\ldots$ stand for extra terms which are proportional in
addition to $sech(\tilde{\lambda} y)^{2,4,6}$. These terms yield
negligible contributions when the $y\to\pm\infty$ integration
limits are taken. This will be especially true when the
$\tilde{\lambda}\to\infty$ limit will be taken, a subtle case,
which is treated separately with a different procedure in
Sec.~\ref{Sec:ActionMatching}. Nevertheless, an explicit
calculation of the integral in this limit will yield the same
$\tilde{\lambda}$ (and $\epsilon$) scaling behavior with a
different numerical pre-factor, which will correspond to the same
late time cosmology on the brane.

Turning to $\Omega_{\dot{\phi}^2}$ and the related
$\Omega_V^{2,3}$ we get:
\begin{equation}
\int d\omega\sqrt{G}\Omega_{\dot{\phi}^2}=\int dy
\frac{3\lambda^2\epsilon}{2\kappa\tilde{\lambda}^2}\,\frac{\sinh^2(\tilde{\lambda}y)}{\cosh^8(\tilde{\lambda}
y)}=\frac{3\lambda^2\epsilon}{2\kappa\tilde{\lambda}^2}\left(\left.\frac{8
\tanh(\tilde{\lambda}y)}{105\tilde{\lambda}}\right|^{y_+\to +\infty}_{y_-\to-\infty}+
...\right)\approx
\frac{8\lambda^2\epsilon}{35\kappa\tilde{\lambda}^3},\label{phidotintegral}
\end{equation}
which implies:
\begin{equation}
\int d\omega\sqrt{G}\Omega_V^2=-\frac{4\tilde{\lambda}^4}{\lambda^2\epsilon^2}
\int d\omega\sqrt{G}\Omega_{\dot{\phi}^2}
\approx-\frac{32\tilde{\lambda}}{35\kappa\epsilon}\label{V2integral}.
\end{equation}
Finally, we turn to $\Omega_V^4$, which seems to have different
localization properties from the other energy density contributions. After integration, we obtain:
\begin{equation}
\int d\omega\sqrt{G}\Omega_V^4=\frac{6\epsilon\lambda^2}{\kappa\tilde{\lambda}^2}\int\frac{\sinh^4(\tilde{\lambda}y)}
{\cosh^8(\tilde{\lambda}y)}dy=\frac{6\epsilon\lambda^2}{\kappa\tilde{\lambda}^2}\left(\left.\frac{2
\tanh(\tilde{\lambda}y)}{35\tilde{\lambda}}\right|^{y_+\to +\infty}_{y_-\to-\infty}+
...\right)\approx\frac{24\lambda^2\epsilon}{35\kappa\tilde{\lambda}^3}\label{V4integral}.
\end{equation}
Summarizing, we see that while the contributions of $\Omega_V^3$
and $\Omega_{\dot{\phi}^2}$ cancel each other, the identical
contributions of $\Omega_{\phi^{\prime 2}}$ and $\Omega_V^1$ add
up to yield an energy density peaked at $y=0$, and scales with
$\tilde{\lambda}$. Together with the contribution coming from
$\Omega_V^2$, this energy density is positive and amounts to
$\Omega_{\tilde{\lambda}^1}^{brane}\simeq64\tilde{\lambda}/(35\kappa\epsilon)$.
As we shall show below the contribution coming from $\Omega_V^4$
 corresponds instead to an instantaneous bulk energy density, which is a direct generalization of the
bulk cosmological constant, $\Lambda_5$, of the maximally 4
symmetric case discussed in \cite{Davidson}. Further insight into the role of these energy densities will
be gained through the procedure described in the next section.
\subsection{The cosmological constant from matching to an
instantaneous RS-like action}\label{Sec:ActionMatching}
To get a better understanding of the localization properties of
$\Omega_{V}^{1,2,3,4}$ and $\Omega_{\dot{\phi}^2,\phi^{\prime
2}}$, in particular in their natural (yet problematic) late time
thin limit, $\tilde{\lambda}\to\infty$, we
first conveniently rewrite the
equations of motion in terms of $f_a(\omega,\tau)$, defined as:
\begin{equation}
e^{f_a(\omega,\tau)}=a(\omega,\tau)\quad\Rightarrow\quad
f_a(\omega,\tau)=-\ln(\sqrt{1+\lambda^2\omega^2\tau^2})\quad\Rightarrow
\quad f_a(y,\tau)=-\ln\cosh (\tilde{\lambda} y)
\label{fadefined},\end{equation} where the last equality is
written on a constant $\tau$ slice according to the coordinate
transformation of Eq.~(\ref{conformalTrans}). Since
$b(\omega,\tau)=\epsilon a(\omega,\tau)$, we have
$H(\omega,\tau)=F(\omega,\tau)=\dot{f_a}$ and
$h(\omega,\tau)=f(\omega,\tau)=f_a^{\prime}$. Thus, it is straightforward to rewrite Eqs.~(\ref{EEij}), (\ref{EEphidot}) and
(\ref{EEphiprime}) in terms of $f_a(\omega,\tau)$ and its
derivatives and subsequently use these equations to rewrite the
action of Eq.~(\ref{action}) in terms of $f_a$.
The resulting action acquires the following form:
\begin{equation}
S=\int
d^5x\sqrt{G}\left(\frac{1}{2}a^{-2}\dot{\phi}^2-\frac{1}{2\epsilon^2}a^{-2}\phi^{\prime
2}-V(\phi)\right)=\int
d^5x\sqrt{G}e^{-2f_a}\frac{3}{\kappa}\left(\frac{f_a^{\prime\prime}}{\epsilon^2}-\ddot{f_a}+\frac{f_a^{\prime
2}}{\epsilon^2}-\dot{f}_a^2\right).\label{Geometrized Action}
\end{equation}
The matching to an instantaneous RS-like action should be
performed in the thin ($\tilde{\lambda}\to\infty$) limit of the
domain wall configuration obtained in
Sec.~\ref{Sec:definiton+solution}. On  constant $\tau$ surfaces
$\Sigma|_{\tau=\tau_0}$ the RS-like action will consist of a bulk
cosmological constant term and a 3-brane cosmological constant, where the brane is located at $y=0$ (or $\omega=0$). To
match we first transform the action of Eq.~(\ref{Geometrized
Action}) to $y$ coordinates and require that:
\begin{eqnarray}
S &=&(\int d\tau)\int
d^3xd\omega\sqrt{G}e^{-2f_a(\omega,\tau)}\frac{3}{\kappa}\left(\frac{f_a^{\prime\prime}}
{\epsilon^2}-\ddot{f_a}+\frac{f_a^{\prime
2}}{\epsilon^2}-\dot{f}_a^2\right)\nonumber\\ &=&(\int d\tau)\int
d^3xdy\sqrt{G}e^{-f_a(y,\tau)}\frac{3}{\kappa}\left(\frac{f_a^{\prime\prime}}{\epsilon^2}+2\frac{f_a^{\prime
2}}{\epsilon^2}-e^{-2f_a}(\ddot{f_a}-\dot{f}_a^2)\right)\nonumber\\
&\,\,\underset{\tilde{\lambda}\to\infty}\longrightarrow \,\,&
(\int e^{f_a}d\tau)\left[\int
dyd^3x\sqrt{g^{\Sigma_0}}(-\Lambda_{bulk}) + \int d^3x
\sqrt{g^{(3)}}(-\Lambda_{3-brane})\right]\,.
\label{ActionMatchingSlinky}
\end{eqnarray}
Using
$\cosh(\tilde{\lambda}y)\underset{\tilde{\lambda}\to\infty}\to
e^{\tilde{\lambda}|y|}/2$ we obtain all relevant quantities on
$\Sigma|_{\tau=\tau_0}$, in the thin ($\tilde{\lambda}\to\infty$)
limit.
\begin{equation}
f_a(y)\underset{\tilde{\lambda}\to\infty}\to -\tilde{\lambda}|y|
\quad \Rightarrow \quad
a(y)\underset{\tilde{\lambda}\to\infty}\longrightarrow
2e^{-\tilde{\lambda}|y|},\label{thinF}
\end{equation}
\begin{equation}
f_a^{\prime}(y)\equiv\frac{df_a(y)}{dy}=-\tilde{\lambda}\tanh(\tilde{\lambda}y)\underset{\tilde{\lambda}\to\infty}
\longrightarrow
-\tilde{\lambda}(\theta(y)-\theta(-y)),\label{thinFprime}\end{equation}
\begin{equation}
f_a^{\prime\prime}(y)\equiv\frac{d^2f_a(y)}{dy^2}=-\tilde{\lambda}^2{\rm
sech}^2(\tilde{\lambda}y)\underset{\tilde{\lambda}\to\infty}\longrightarrow-2\tilde{\lambda}\delta(y).
\label{thinFprimeprime}\end{equation}
We immediately realize that the $\tilde{\lambda}\to\infty$ limit
of $f_a^{\prime\prime}(y)$ in Eq.~(\ref{ActionMatchingSlinky}),
will generate a brane localized energy density proportional to
$\tilde{\lambda}$,
$\Lambda_{3-brane}^{f_a^{\prime\prime}}=6\tilde{\lambda}/(\kappa\epsilon)$,
which agrees with the previous results obtained by analyzing the
various terms in $T_{00}$ in Sec.~\ref{Sec:LambdaSlinky}. In
particular the brane cosmological constant,
$\Lambda_{3-brane}^{f_a^{\prime\prime}}$, is related to the
contributions of the  $\Omega_{\phi^{\prime 2}}$ and $\Omega_V^2$
densities of Eqs.~(\ref{DensitiesDefprime}) and
(\ref{DensitiesDefPotential}) by:
\begin{equation}
\Omega_{\phi^{'2}}+\Omega_V^1
\underset{\tilde{\lambda}\to\infty}\longrightarrow
T_{00}(\Lambda_{3-brane}^{f_a^{\prime\prime}})=e^{2f_a}\delta(y)\Lambda_{3-brane}^{f_a^{\prime\prime}}
\label{DensitiesActionCorespondence}
\end{equation}

The contribution coming from $f_a^{\prime 2}(y)$ will instead correspond to
the aforementioned generalization of the bulk cosmological
constant, $\Lambda_5$, appearing in \cite{Davidson} and
Sec.\ref{Sec:LambdaM4}. These contributions are $y$ independent in
the $\tilde{\lambda}\to\infty$ limit and will be specified later.
 \noindent To obtain the explicit form for
the $\dot{f}_a$ and $\ddot{f}_a$ terms we act  with derivatives
with respect to $\tau$ on $f_a(\omega,\tau)$, since the coordinate
transformation of Eq.~(\ref{conformalTrans}) is defined only on
constant $\tau$ surfaces. Once the derivatives have been obtained
we express them in terms of $y$ on $\Sigma|_{\tau=\tau_0}$ which
results in:
\begin{equation}
\dot{f}_a\equiv\frac{df_a(\omega,\tau)}{d\tau}\vert_{\Sigma|_{\tau=\tau_0}}
=-\frac{\lambda\tilde{\lambda}\omega^2}{1+{\tilde{\lambda}}^2\omega^2}
=-\frac{\lambda\sinh^2(\tilde{\lambda}y)}{\tilde{\lambda}
\cosh^2(\tilde{\lambda}y)}\underset{\tilde{\lambda}\to\infty}\longrightarrow
-\frac{\lambda}{\tilde{\lambda}}\left(\theta(y)-\theta(-y)\right)^2=-\frac{\lambda}{\tilde{\lambda}}\, .
\label{thinFdot}\end{equation} This contribution also corresponds to a $y$
independent (totally delocalized in $y$) energy density in the
$\tilde{\lambda}\to\infty$ limit. Finally, we inspect the
$\tilde{\lambda}\to\infty$ limit of $\ddot{f}_a$ in an analogous
way to $\dot{f}_a$ and obtain:
\begin{eqnarray}
\ddot{f}_a\equiv\frac{d^2f_a(\omega,\tau)}{d\tau^2}\vert_{\Sigma|_{\tau=\tau_0}}=\frac{2{\tilde{\lambda}}^2\lambda^2\omega^4}
{(1+{\tilde{\lambda}}^2\omega^2)^2}-\frac{\lambda^2\omega^2}{1+{\tilde{\lambda}}^2\omega^2}\nonumber&=&\frac{\lambda^2}
{\tilde{\lambda}^2}\left(2\tanh^4(\tilde{\lambda}
y)-\tanh^2(\tilde{\lambda}
y)\right)\\\underset{\tilde{\lambda}\to\infty}\longrightarrow\frac{\lambda^2}
{\tilde{\lambda}^2}\left(2(\theta(y)-\theta(-y))^4-(\theta(y)-\theta(-y))^2\right)=\frac{\lambda^2}
{\tilde{\lambda}^2}\label{thinFdotdot}
\end{eqnarray}
From Eqs.~(\ref{ActionMatchingSlinky}), (\ref{thinFdot})
and~(\ref{thinFdotdot}) we realize that the
$\tanh(\tilde{\lambda}y)^4$ terms in $\dot{f_a}^2$ and
$\ddot{f}_a$ are added to each other, leaving us with both
$\tanh^2(\tilde{\lambda}y)$ and $\tanh^4(\tilde{\lambda}y)$ terms,
which will together contribute to a $y$ dependent bulk energy
density. Their contribution enters with a conformal prefactor
$e^{-2f_a}$, implying that the associated bulk energy density,
$\Omega_{\dot{f}_a^2,\ddot{f}_a}$ behaves as
$\cosh^2(\tilde{\lambda}y)\tanh(\tilde{\lambda}y)^{(2,4)}$.
However, since the metric is strongly peaked at $y=0$ in the
$\tilde{\lambda}\to\infty$ limit and we care about the
contribution of $\Omega_{\dot{f}_a^2,\ddot{f}_a}$ to the induced
instantaneous brane cosmological constant, the  $e^{-2f_a}$
conformal prefactor does not modify the step function limits of
Eqs.~(\ref{thinFdot}) and (\ref{thinFdotdot}) relevant for our purpose.
Notice also that if we
integrate over $y$, the total contribution of these energy
densities remains finite.
The associated bulk
contribution to the brane induced cosmological constant amounts to
$\Lambda_{bulk}^{\dot{f}_a,\ddot{f}_a}=(6/\kappa)\lambda^2/\tilde{\lambda}^2$
in the $\tilde{\lambda}\to\infty$ limit. This result
 corresponds to the contribution of
$\Omega_V^4=(6/\kappa)(\lambda^2/\tilde{\lambda}^2)\tanh^4(\tilde{\lambda}y)$
in Eq.~(\ref{DensitiesDefPotential}).

Finally, from Eqs.~(\ref{ActionMatchingSlinky}) and
(\ref{thinFprime}) the contribution of the $f_a^{\prime 2}$ term
to the bulk energy density (on $\Sigma_{\tau=\tau_0}$),
$\Lambda^{f_a^\prime}_{bulk}=-(6/\kappa)\tilde{\lambda}^2/\epsilon^2$
corresponds to the contribution of $\Omega_V^2$,
which is thus interpreted as a bulk energy term.
\section{Simplest Cosmological
scenarios}\label{Sec:CosmologicalScenarios}
The above analysis
tells us that in the thin ($\tilde{\lambda}\to\infty$) limit the
slinky configuration introduced in
Sec.~\ref{Sec:definiton+solution} is equivalent to a brane in a
$AdS_5$-like\footnote{$AdS_5$-like means that constant $\tau$ slices look like a slice of $AdS_5$ with different conformal factor. The latter smoothly varies with time.} bulk characterized by the warp factor
$a(y)=e^{-\tilde{\lambda}|y|}$. We have found that the bulk and brane
cosmological constants are given by:
\begin{equation}
\Lambda_5^{sl.}=\frac{6}{\kappa}\left(\frac{\lambda^2}{\tilde{\lambda}^2}-\frac{\tilde{\lambda}^2}{\epsilon^2}\right)
\qquad
\Lambda_b^{sl.}=\frac{6}{\kappa}\frac{\tilde{\lambda}}{\epsilon}
\label{ThinLimitConstants}.\end{equation}
From the above equation, we immediately realize that the RS fine
tuning relation is asymptotically satisfied on every constant
$\tau$ slice, in the limit $\tilde{\lambda}\to\infty$ (;large
times) thus supplementing us with a static brane. This happens for
any value of the parameter $\epsilon$. Also notice that, unlike
the solution in \cite{Giovannini}, $\epsilon=1$ does not
correspond to a free solution. At any finite $\tilde{\lambda}$, we
have an induced brane dark energy density given by:
\begin{equation}
(\Lambda_{b}^{ind.})_{slinky}=\kappa\frac{(\Lambda_b^{sl.})^2}{6}+\Lambda_5^{sl.}=\frac{6\tilde{\lambda}^2}{\kappa}
\left(\frac{1}{\epsilon^2}-\frac{1}{\epsilon^2}\right)+\frac{6\lambda^2}{\kappa \tilde{\lambda}^2}
=\frac{6\lambda^2}{\kappa \tilde{\lambda}^2}\,.\label{InducedSlinkyDensity}\end{equation}
Notice that  the brane cosmological constant, $\Lambda_b^{sl.}$
enters quadratically in the RS fine tuning relation. This is in
exact correspondence to the well known analysis of fundamental
brane cosmology by Deffayet {\it et al}.
\cite{BraneCosmology1:Binetruy:1999ut,BraneCosmology5:Binetruy:1999hy},
as follows. In the general time-dependent fundamental brane case,
one solves the 5-dimensional Einstein equations by imposing the
Israel matching conditions relating the discontinuities in the
derivatives of the metric components across $y = 0$, to delta
distribution sources (see \cite{BraneCosmology1:Binetruy:1999ut}).
It turns out that the behaviors of the brane sources and the
metric components evaluated at $y=0$ are independent of the metric
solutions in the bulk, and obey a modified Friedmann equation
where the brane energy density enters quadratically.

Thus, the late time (large $\tilde{\lambda}$) cosmology of the
simplest slinky configuration reported in
Sec.~\ref{Sec:definiton+solution}, in the absence of additional
sources, is simply driven by a dark energy density
$(\Lambda_{b}^{ind.})_{slinky}$, which will consist only of the
bulk contributions generated by the \footnotesize{$\dot{f}_a^2$}
and \footnotesize{$\ddot{f}_a$} \normalsize terms in the action of
Eq.~(\ref{ActionMatchingSlinky}), which decay like $1/\tau^2$ for
large values of $\tau$.

To conclude the analysis of the late time ($\tau>>1$) cosmology on
the brane we write (and solve) the effective Friedmann equation
for the evolution of the scale factor of the instantaneous
3-brane:
\begin{equation}
H_b(\tau)^2=\kappa\Lambda_b^{ind.}=\frac{6}{\tau^2} \Rightarrow
a_b(\tau)=e^{\int H_b(\tau)d\tau}=C_b\,\tau^{\sqrt{6}}
,\label{FriedmannSlinky}\end{equation} where $H_b(\tau)$ is the
brane Hubble constant and $C_b$ is an integration constant. Thus,
the late time brane is accelerating with a power law for the scale
factor, corresponding to a deceleration parameter,
$q=-a_b\,\ddot{a_b}/\dot{a}_b^2=1/\sqrt{6}-1\simeq -0.6$, which is
rather plausible if we simply treat $\tau$ as the cosmological
time\footnote{In the $\Lambda$CDM model, the observed dark matter
and dark energy density parameters are \cite{WMAP2010}
$\Omega_M\simeq 0.3$ and $\Omega_\Lambda \simeq 0.7$,
respectively. Together, they yield a deceleration parameter,
$q_{\Lambda CDM}=\Omega_M/2-\Omega_\Lambda\simeq-0.55$.}.

 The addition of (a-priori) brane localized sources
(matter, radiation), will obviously enable various modifications
of the  brane cosmological evolution, all of which will be
analogous to the analysis of
\cite{BraneCosmology1:Binetruy:1999ut,BraneCosmology5:Binetruy:1999hy}.
Needless to say, the latter possibility is highly implausible due
to the large element of arbitrariness or the lack of underlying
dynamics.

Finally, we comment on the early time cosmological interpretation
of the solution reported in Sec.~\ref{Sec:definiton+solution}.
Since at early times ($\tilde{\lambda}\to 0$) the kink
configuration and warp factor are nearly flat ($\phi\to 0$ and
$a(\omega,\tau)\to 1$), we should expect that the average of the
large early time bulk energy densities,
$\Omega_{\dot{f}_a,\ddot{f}_a}$, will dictate the cosmological
evolution of the 4D geometry obtained by integrating out $\omega$
(or $y$).

The main problem with this early time picture is the absence of a
brane, or a specific region in the extra dimension, corresponding
to the 4D DW brane universe we wish to obtain in the first place.
It is at this point that we take advantage of the fact that the
solution reported in Sec.~\ref{Sec:definiton+solution} is
symmetric in $\omega$ and $\tau$. By performing the same analysis
of Sec.~\ref{Sec:ActionMatching} on constant $\omega$ slices,
$\Sigma|_{\omega=\omega_0}$, we realize that in the limit
$\omega\to\infty$ the scalar field, $\phi$, becomes a ``brane in
time" ($\tau$-brane) localized at $\tau=0$. Similarly, the
$\ddot{f}_a$ term in the action (Eq.~(\ref{Geometrized Action})),
will become a delta distribution in time for $\omega\to\infty$ ,
in analogy with Eq.~(\ref{thinFprimeprime}) and will thus
correspond to a brane localized energy density.

Given the above, the hypersurface $\Sigma|_{\tau=0}$ will
correspond to the center of a thick ``$\tau$-brane", with a brane
localized energy density that diverges as $\omega\to\infty$ -
thus providing  an element of creation, which occurs
smoothly for any finite value of $\omega$, in the spirit of the
Hartle-Hawking no boundary proposal.

To conclude, we find the cosmological interpretation of the
slinky solution introduced  in Sec.~\ref{Sec:definiton+solution} and studied
above to be plausible and problematic at the same time.
It is plausible, because an extremely simple 5D
setup with a single scalar field is able to generate a DW brane
universe with an early and late time accelerating phases, thus
addressing directly the dark energy paradigm.

The main problem with the same  interpretation stems from the
subtle identification of early time dynamics and the interpolation
between the late time and early time regimes. The most natural
explanation, which seems to suggest itself, is that our DW brane
universe is localized towards $\omega\to\infty$ as $\tau\to 0$
(which can also be perceived as its creation at this point in the
extra dimension) and evolves to an infinitely thin brane at
$\omega=0$ for $\tau\to\infty$, thus corresponding to a different
(non trivial) watershed than the one we naively started with,
namely the line $\omega=0$. To realize such a possibility one must
find a new solution in the intermediate time regime, which is able
to interpolate between the above early and late time behaviors.
Simultaneously, this interpolating solution should satisfy the
$(05)$ constraint (Eq.~(\ref{EE05})), to balance the DW-bulk
energy transfer in intermediate times (Eq.~(\ref{05Violation})).
Further study of this possibility and the search for alternative
solutions will be the subject of future publications.


\section{Conclusions}\label{Sec:Conclusions}

In this paper we have proposed a novel type of time dependent DW
configurations in a warped 5D space time. The purpose is to find a
dynamical realization for braneworld cosmology. After discussing
the initial symmetry of the time $t$ and some extra spatial
dimension $y$ prior to the dynamics generating the DW brane
universe, we have introduced the notion of slinky configurations.
We require that such configurations satisfy the most obvious
properties for a cosmologically plausible time dependent DW brane
setup, namely an element of creation, a finite inflationary period
and late time acceleration.

\noindent The
simplest configuration we were able to find was based on a time ``proportional'' ansatz for the metric warp factors, corresponding to a
conformally flat 5D metric. This configuration has most of the desired features, however it introduces an explicitly coordinate dependent scalar potential. As part of future steps in this direction, it will be important to clarify the role and implications of properties such as non factorizability in $(t,y)$ of the solutions and how these properties are related to the possible loss of 5D covariance.

\noindent The cosmological
evolution of the proposed DW brane configuration was studied by constructing instantaneous
energy densities associated with the DW scalar and its supporting
potential on constant time ($\tau$) slices, where the potential can be
written in a covariant form. This procedure allowed to identify bulk and brane induced energy densities (cosmological constants).
The resulting cosmology was shown to include
a smooth element of creation (in the spirit of the Hartle-Hawking no
boundary proposal \cite{Hartle-Hawking}), a
naturally terminating inflationary period and a late time
acceleration phase.

\noindent An underlying dynamics that accounts for more detailed
cosmological features would in principle require the addition of matter
and gauge fields, rendering the problem even more difficult to solve.
No satisfactory solution has been found up to date. Another, simplifying avenue
is to study cosmological perturbations on a given slinky
configuration. We plan to search for new more satisfactory solutions in the context of slinky configurations and follow the proposed avenues in order to probe their cosmological implications.

\noindent The
long term purpose of this study is to find a cosmologically
plausible and fully dynamical DW brane setup, which unifies cosmological and particle physics aspects, and can be
used for particle physics in the spirit of
\cite{DamienTime}. A configuration able to satisfy simultaneously
the constraints coming from both cosmological observations and
collider physics experiments is certainly highly desirable;
it is also appealing the possibility that its building blocks
require nothing more than General Relativity and Quantum Field Theory in 5D.


\bibliography{SlinkyThoughts}
\bibliographystyle{utphys}

\end{document}